\documentclass[10pt,twoside]{hsqcd}
\usepackage{epsf,amsmath}
\usepackage{graphicx}

\def\prd#1#2#3{    {\it Phys. Rev. }{\bf D#1} (19#2) #3}

\def\eq#1{{Eq.~(\ref{#1})}}   

\newcommand{\beq}{\begin{equation}}   
\newcommand{\eeq}{\end{equation}} 

\begin{document}

\title{HIGH ENERGY EVOLUTION  BEYOND \\
THE BALITSKY - KOVCHEGOV EQUATION 
\thanks{The author was honored to be a DESY Theory Group fellow
when giving this talk}
}

\author{M.~LUBLINSKY \\
Department of Physics, U-3046, 
University of Connecticut \\
2152 Hillside Rd., Storrs, CT-06269, USA \\
E-mail: lublinm@mail.desy.de }

\maketitle

\begin{abstract}
\noindent We present a short review of  recent theoretical  
activity which is an attempt to learn about low $x$ evolution
 beyond the Balitsky - Kovchegov equation.  
\end{abstract}



\markboth{\large \sl M. Lublinsky  \hspace*{2cm} HSQCD 2004}
{\large \sl \hspace*{1cm} TEMPLATE FOR THE HSQCD'04 PROCEEDINGS}

\section{Balitsky-Kovchegov equation}
The Balitsky - Kovchegov equation (BK) \cite{GLR,BA,KO,Braun,ILM}
 is the best presently available 
tool to study saturation phenomena at high energies. Contrary to many 
models the BK has solid grounds in perturbative QCD.  The equation reads   
\beq 
\label{EQ}   
 \frac{d\,N({\mathbf{x_{01}}},y;b)}{d\,y}\,= 
\,\frac{\bar\alpha_s}{2} \,  
 \int\, d^2 {\mathbf{x_{2}}}    
\frac{{\mathbf{x^2_{01}}}}{{\mathbf{x^2_{02}}}\,   
{\mathbf{x^2_{12}}}} \,\,\times \,\,\,\,\,\,\,\,\,\,\,\,\,\,\,\,\,\,\,\,
\,\,\,\,\,\,\,\,\,\,\,\,\,\,\,\,\,\,\,\,
\,\,\,\,\,\,\,\,\,\,\,\,\,\,\,\,\,\,\,\,\,\,\,\,\,\,\,\,\,\,\,\,\,\,\,\,\,\,\,\,
\,\,\,\,\,\,\,\,\,\,\,\,\,\,\,\,\,\,\,\,
\eeq
$$   
\left(2\,  N({\mathbf{x_{02}}},y;{ \mathbf{ b-   
\frac{1}{2}   
x_{12}}})\,- \,N({\mathbf{x_{01}}},y;\mathbf{b})\,-\, N({\mathbf{x_{02}}},y;{ \mathbf{ b -   
\frac{1}{2}   
x_{12}}})  N({\mathbf{x_{12}}},y;{ \mathbf{ b- \frac{1}{2}   
x_{02}}})\right)   
$$   
The function $N(r_{\perp},y; b)$  
stands for imaginary part of the amplitude for a dipole of size   
$r_{\perp}=\mathbf{x_{10}}$ 
elastically scattered at the impact parameter $b$.   
 \eq{EQ} has a  simple meaning:  the dipole of size   
$\mathbf{x_{10}}$ decays in two dipoles of  sizes   
$\mathbf{x_{12}}$ and $\mathbf{x_{02}}$   
 with the decay probability   
  $\frac{\mathbf{x^2_{01}}}{\mathbf{x^2_{02}}\,\mathbf{x^2_{12}}}$.   
 These two dipoles   
 then interact with the target. The non-linear term  takes into account   
a  simultaneous interaction of two produced dipoles with the   
target. The linear part of \eq{EQ} is the LO BFKL equation,   
which describes the evolution of the multiplicity of the fixed   
size color dipoles   with respect to the rapidity (energy) $y$. 
For the discussion  
below we introduce a short notation for the BK: 
$ 
\frac{d\,N}{d\, y}\,\,=\,\alpha_s\,\,Ker\, \,\otimes\,\,(\,N\,\,- 
\,\,N\,\,N)\,.
$

The theoretical success associated with the BK is based on 
the following facts: 
$\bullet$
The BK is based on the correct high energy dynamics which is taken into 
account via the LO BFKL evolution kernel.  
$\bullet$
The BK  restores the s-channel unitarity of partial waves  
 (fixed impact parameter) which is badly violated by the linear  
BFKL evolution.  
$\bullet$ The BK is believed to describe gluon saturation, 
a phenomenon expected at high 
energies.  
$\bullet$ The BK resolves the infrared diffusion problem associated 
with the linear BFKL evolution. This means the equation is much more stable 
with respect to possible corrections coming from the nonperturbative domain. 
$\bullet$
 The BK has met with phenomenological successes when confronted against 
DIS data from HERA \cite{GLLM,GLLMN,Iancu}.

The BK is not exact and has been derived in several approximations. 
$\bullet$
The  LO BFKL kernel is obtained in the leading  
soft gluon emission approximation  and at fixed $\alpha_s$. 
 $\bullet$ The Large $N_c$ limit is used in order to express
 the nonlinear term 
as a product of two functions $N$. This limit is in the foundation of 
the color  dipole picture. To a large extent the large $N_c$ limit is 
equivalent to a  mean field theory without dipole correlations. 
 $\bullet$ The BK assumes no target correlations. Contrary to the large 
$N_c$ limit which is a controllable approximation within perturbative QCD, the 
absence of target correlations is of pure nonperturbative nature. This  
assumption has grounds for asymptotically  heavy nuclei, but it is likely 
not  be valid for proton  or realistic nucleus target.

There are several quite serious  theoretical problems which need to  
be resolved in the future. 
$\bullet$
The BK is not symmetric with respect to target and projectile. 
While the latter is assumed to be small and perturbative, the former 
is treated as a large nonperturbative object. 
The fan structure of the diagrams 
summed by the BK violates the t-channel unitarity. A first step towards 
restoration of the t-channel unitarity would be an inclusion of 
Pomeron loops.  

 $\bullet$
 Though the BK respects the s-channel unitarity  
\footnote{There was a recent claim of Mueller and Shoshi \cite{MS} that the  
$s$-channel unitarity is in fact violated during the evolution.}  
it violates the Froissart bound for  energy behavior of the  
 total cross section. One needs gluon saturation and confinement in order 
to respect the Froissart bound. On one hand, the BK provides the  
gluon saturation at fixed  impact parameter. On the other hand, 
being purely perturbative it cannot generate a mass gap needed to  
ensure a fast convergence of the integration over the impact parameter $b$. 
Because of this problem, up to now all the phenomenological applications 
of the BK were based on model assumptions regarding the $b$-dependence. 
It is always assumed that the $b$-dependence factorizes and in practice 
the BK is usually solved without any trace of $b$.  An attempt to go beyond  
this approximation was reported in  \cite{BKb}. 
 
$\bullet$
 It is very desirable to go beyond the BK and relax all underlying  
assumptions outlined above. The high order corrections are most needed. In  
particular it is important to learn how to include running of $\alpha_s$  
though in the phenomenological applications the running of $\alpha_s$ has  
been usually  implemented.  
 
$\bullet$
 The LO BFKL kernel does not have the correct short distance limit 
 responsible for the Bjorken scaling violation. As a result the BK does  
not naturally match with the DGLAP equation. Though several approaches  
for unification of the BK and DGLAP equations were proposed  
\cite{LGLM,GLLM,KMS}, the methods are not fully developed. All approaches 
deal with low $x$ and gluon sector only. We would like to have a unified  
evolution scheme both for small and large $x$ with quarks equally treated.

\section{Beyond the BK}  
We now present a short (not complete) review of a recent theoretical  
activity which is an attempt to go beyond the BK.   
 
$\bullet$ Though the BFKL kernel is known at next-to-leading order, 
a nonlinear equation at NLO has not been derived yet.  
The authors of \cite{BB} have been able to compute a single 
NLO contribution which has maximal nonlinearity, namely the $N^3$ term:   
\beq\label{BB} 
\frac{d\,N}{d\, y}\,\,=\,\alpha_s\,Ker\, \,\otimes\,\,(\,N\,\,- 
\,\,N\,\,N)\,\,-\,\, 
{\alpha_s^2\,\tilde{Ker}\,\,\otimes\,\,N\,N\,N}\,. 
\eeq 
In \cite{DT}  the NLO BFKL at presence of a 
saturation boundary was considered. 
The results show a decrease in the saturation scale  
growth as a function of rapidity towards the value $\lambda\,\simeq \,0.3$ 
observed experimentally.  
    
In \cite{CLSV} a study of the BK at 
presence of a rapidity veto was made. 
Rapidity veto means that no parton emissions 
are allowed which are separated in rapidity by less than the veto $\eta$. 
At high energies the method of rapidity veto is known to mimic  higher 
order corrections. The application of the method to the BK equation 
makes it nonlocal in rapidity   
$$ 
\frac{d\,N(y)}{d\, y}\,\,=\,\alpha_s\,Ker\, \,\otimes\,\,(\,N(y\,{ -\,\eta}) 
\,\,-\,\,N(y\,{-\,\eta})\,\,N(y\,{ -\,\eta})) \,.
$$ 
The veto delays saturation in accord with the expectations associated 
with the next-to-leading order corrections. If the veto is put on top of the
BK equation with running $\alpha_s$ then the effect of additional 
NLO corrections
is significantly reduced. This observation gives support to phenomenological
studies of Refs.   \cite{GLLM,KK}. 
\\ 
Another approach  to  partially include NLO corrections into BK
equation is to implement in its linear term   a unified
BFKL-DGLAP framework developed in \cite{KMS,KK}.

$\bullet$  The $N_c$ corrections can be accounted for through 
JIMWLK functional equation \cite{ELTHEORY},  
which is equivalent to the   
Balitsky`s infinite hierarchy of equations \cite{BA}.   
Introducing $N$ as a target expectation value 
of a certain operator (product of two Wilson lines), 
$N\,\,\equiv\,\,\langle\,W\,\rangle_{target}$, the first couple 
of equations are 
\beq 
 \hspace{1cm} 
\frac{d\,\langle\,W\,\rangle}{d\, y}\,\,=\,\alpha_s\,Ker \,\otimes\,(\, 
\langle\,W\,\rangle\,\,- 
\,\,\langle\,W\,W\,\rangle). 
\eeq  
\beq  
\frac{d\,\langle\,W\,W\,\rangle}{d\, y}\,=\alpha_s 
\,Ker\, \otimes\,( 
\langle\,W\,W\,\rangle\,- 
\,\langle\,W\,W\,W\,\rangle). 
\eeq 
The large $N_c$ limit and the absence of the target corellations 
used by Kovchegov \cite{KO} is equivalent to a  
mean field approximation which allows to express a correlator of a product 
as a product of correlators: $\langle\,W\,W\,\rangle\,\, 
=\,\,\langle\,W\,\rangle\,\langle\,W\,\rangle\,\,=\,\,N\,N\,;  
\hspace{1.cm}N_c\,\rightarrow\,\infty$. Thus the first equation of the 
Balitky`s chain closes to the BK.   
 
A first 
numerical solution of the JIMWLK equation was reported in \cite{Weigert}. 
They do not find any qualitative 
deviation from solutions of the BK. The $N_c$ corrections were found 
to be at a level of few pecents.
 
The authors of \cite{BLV} have considered 
  $N_c$ corrections to the triple Pomeron vertex: 
$$ \hspace{1cm} 
\frac{d\,N}{d\, y}\,\,=\,\alpha_s\,\,Ker\, \,\otimes\,\,(\,N\,\,- 
\,\,N\,\,N\,\,-\,\,{ \frac{1}{N_c^2}\,n})\, 
$$ 
An additional equation for $n$ was proposedin \cite{BLV}.  
 
$\bullet$ For proton and realistic (not very dense) nucleus targets a 
systematic approach towards  inclusion of target  
correlations has been developed in \cite{LL}.    
Target correlations can be introduced via certain  
linear functional differential equation.  In general, this linear 
functional equation cannot be reformulated as a nonliner equation. 
However, in a particular case when all $n$-dipole correlations can be 
accounted for by a single correlation parameter, the equation can be 
brought to a modified version of the BK:   
\beq\label{LLcor}  
\frac{d\,N}{d\, y}\,=\,\alpha_s\,Ker\, \otimes\,(\,N\,- 
\,\kappa \,N\,N), 
\eeq 
with $\kappa \ge 1$ being the correlation parameter to be found
from a model for the target.

$\bullet$ 
Inclusion of Pomeron loops is the first step towards restoration of the 
$t$-channel unitarity. Iancu and Mueller \cite{IM} has recently considered
rare fluctuations which were interpreted in
 \cite{KL} as pomeron loop contributions.
Unfortunately, it looks like 
 contributions of the  pomeron loops cannot be incorporated in a framework 
of a single equation. They  modify the asymtotic behavior of 
the amplitude $N$ in the deep saturation limit:  
$$ N(Y)\,\,=\,\,1\,\,-\,\,e^{\,-\,c\,(Y\,-\,Y_0)^2}\,; \hspace{0.8cm}  
Y\,\rightarrow\,\infty \hspace{1.5cm} c\,=\,2\,\bar\alpha_s\, 
\hspace{1.5cm}  BK 
$$  
$$ 
N(Y)\,\,=\,\,1\,\,-\,\,e^{\,-\,{ 1/2}\,c\,(Y\,-\,Y_0)^2}\,;  
\hspace{0.8cm}  
Y\,\rightarrow\,\infty \hspace{2.5cm}  Pom\,\, Loops 
$$

$\bullet$ It is claimed that the BK sums all possible contributions which are 
not suppressed either by $\alpha_s$ or $N_c$. For example, the cubic term  
which is in Eq. \ref{BB}  appears at next-to-leading $\alpha_s$ order only. 
In particular it is implied that all multi-pomeron exchanges and multipomeron 
vertices are either absorbed by the triple pomeron vertex of the BK or  
suppressed.  It was argued in \cite{LL} that this 
might be false. They argue that in addition to a possibility for a pomeron 
to split into two, there exists a local in rapidity process of multi-pomeron 
exchange. After these contributions were resummed in the eikonal 
approximation, 
a new modification of the BK was proposed: 
\beq\label{LL} 
\frac{d\,N}{d\, y}\,\,=\,\,(1\,\,-\,\,N)\,\alpha_s\,\,Ker\, \, 
\otimes\,\,(\,N\,\,-\,\,N\,\,N) 
\eeq

$\bullet$ Outlook.  It is essential for the future phenomenological 
studies to eliminate the model dependent treatments of the impact parameter. 
Though the BK has been solved numerically with the full $b$-dependence traced 
\cite{BKb}, the results are not yet suitable for phemnomelogical  
applications.

A further study of the relation between the dipole picture vs.  
traditional diagramatics based on the $s$-channel unitarity is needed. 
In particular, it is not clear if the dipole picture survies at NLO.  
In general there is a quest for a {\bf simple} effective Reggeon field  
theory in QCD.

\end{document}